\documentclass{ifacconf}

\usepackage{natbib}        % required for bibliography

\usepackage{graphicx} 

\graphicspath{{./figures/}}
\DeclareGraphicsExtensions{.pdf,.png,.eps,.ps}

\usepackage{amsmath,amssymb} % amssymb internally loads amsfonts
\usepackage{bbm}
\usepackage{enumerate}
\usepackage{float}
\usepackage{color}
\usepackage[Algorithm]{algorithm}
\usepackage{algpseudocode}
\algdef{SE}[DOWHILE]{Do}{doWhile}{\algorithmicdo}[1]{\algorithmicwhile\ #1}%

\usepackage{subfigure}

\newcommand{\node}{N}
\newcommand{\link}{L}
\newcommand{\linkset}{\mathcal{L}}
\newcommand{\junction}{J}
\newcommand{\phases}{\mathcal{P}}
\newcommand{\nodeset}{\mathcal{N}}
\newcommand{\junctionset}{\mathcal{J}}

\newcommand{\expect}{\mathbb{E}}

\newcommand{\inputnodes}{\mathcal{I}}
\newcommand{\outputnodes}{\mathcal{O}}

\begin{document}

\begin{frontmatter}

\title{Back-pressure traffic signal control \\ with unknown routing rates} 

\author[First]{Jean Gregoire} 
\author[Second]{Emilio Frazzoli} 
\author[Third]{Arnaud de La Fortelle}
\author[Fourth]{Tichakorn Wongpiromsarn}

\address[First]{Mines ParisTech, Paris, France \\jean.gregoire@mines-paristech.fr.}
\address[Second]{Massachusetts Institute of Technology, Boston, USA \\frazzoli@mit.edu}
\address[Third]{Mines ParisTech, Paris, France; Inria Paris-Rocquencourt, France \\arnaud.de\_la\_fortelle@mines-paristech.fr}
\address[Fourth]{Thailand Center of Excellence for Life Sciences, Thailand \\tichakorn@tcels.or.th}

\begin{abstract}               
The control of a network of signalized intersections is considered. Previous works proposed a feedback control belonging to the family of the so-called back-pressure controls that ensures provably maximum stability given pre-specified routing probabilities. However, this optimal back-pressure controller (BP*) requires routing rates and a measure of the number of vehicles queuing at a node for each possible routing decision. It is an idealistic assumption for our application since vehicles (going straight, turning left/right) are all gathered in the same lane apart from the proximity of the intersection and cameras can only give estimations of the aggregated queue length. In this paper, we present a back-pressure traffic signal controller (BP) that does not require routing rates, it requires only aggregated queue lengths estimation (without direction information) and loop detectors at the stop line for each possible direction. A theoretical result on the Lyapunov drift in heavy load conditions under BP control is provided and tends to indicate that BP should have good stability properties. Simulations confirm this and show that BP stabilizes the queuing network in a significant part of the capacity region. 
\end{abstract}


\end{frontmatter}
%===============================================================================

\section{Introduction}

\begin{figure}[ht]
\centering
\includegraphics[width=0.6\linewidth]{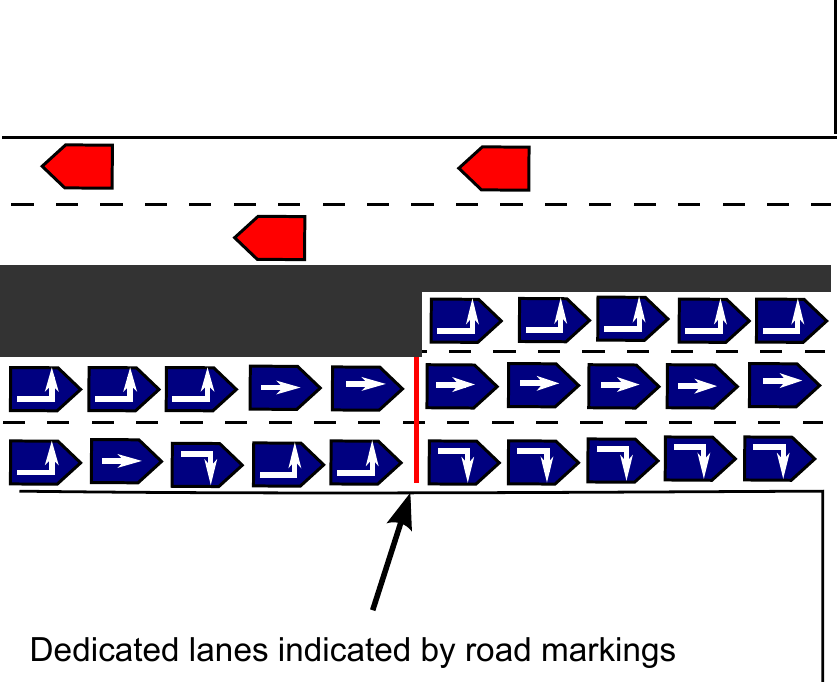}\hfill
\caption{Dedicated lanes for turning vehicles. The dedicated lanes are indicated by road markings when vehicles approach the intersection. Apart from the proximity of the intersection, vehicles are all gathered.}
\label{fig-vehicles-gathered}
\end{figure}

In today's metropolitan transportation networks, traffic is regulated by traffic light signals which alternate the right-of-way of users (e.g., cars, public transport, pedestrians). Congestion is a major problem resulting in a loss of utility for all users due to delayed travel times over the network~\cite{Shepherd1992}. That is why it is of high interest to find a control policy that can stabilize a network of signalized intersections under the largest possible arrival rates. Under traffic light control, a particular set of feasible simultaneous movements, called a phase, is decided for a period of time ~\cite{Papageorgiou2003}. Controlling a traffic light consists of designing rules to decide which phase to apply over time. Pre-timed policies activate phases according to a time-periodic pre-defined schedule, and the signal settings can be fixed by optimization, assuming  within-day static demand \cite{cascetta2006models, miller1963settings, gartner1975optimization}. They are not efficient under changing arrival rates which require adaptive control. Many major cities currently employ adaptive traffic signal control systems including SCOOT~\cite{Hunt1982}, SCATS~\cite{Lowrie1990}, PRODYN~\cite{Henry1984}, RHODES~\cite{Mirchandani2001}, OPAC~\cite{Gartner1983} or TUC~\cite{Diakaki2002}. These systems update some control variables of a configurable pre-timed policy on middle term, based on traffic measures. Control variables may include phases, splits, cycle times and offsets~\cite{Papageorgiou2003}. More recently, feedback control algorithms that ensure maximum stability have been proposed both under deterministic arrivals~\cite{Varaiya2013}, and stochastic arrivals~\cite{Varaiya2009,Wongpiromsarn2012}. These algorithms are based on the so-called back-pressure control presented in the seminal paper~\cite{Tassiulas1992} for applications in wireless communication networks and require real-time queues estimation. An optimal back-pressure traffic signal controller (BP*) is presented in \cite{Wongpiromsarn2012} and \cite{Varaiya2009}. They are defined under different modelling assumptions but they are algorithmically equivalent. The key benefit of back-pressure control is that it can be completely distributed over intersections, i.e., it requires only local information and it is of $\mathcal{O}(1)$ complexity. However, the strong assumptions of the model in \cite{Varaiya2009} (and also implicitly in \cite{Wongpiromsarn2012}) is that controllers require routing rates and a measure of the number of vehicles queuing at every node of the network for each possible routing decision. However, in reality, apart from the proximity of the intersection, vehicles (going straight, turning left, turning right, etc.) are all gathered, and it is difficult to estimate the number of vehicles queuing for each direction (see Figure \ref{fig-vehicles-gathered}). Cameras can give good estimations of the total number of vehicles queuing at a given node, but not the direction of vehicles. However, it is feasible to detect if there are some vehicles (or no vehicle) that want to go to a given destination, if we assume the existence of dedicated lanes from the proximity of the intersection with loop detectors at the stop line. 

The back-pressure control proposed in this paper(BP) requires such loop detectors and an estimation of the total number of vehicles queuing at each node (gathering all possible directions). It does not assume any knowledge of routing rates. We evaluate the performance of BP with regards to the optimal BP* control. The contribution of the paper is to provide a back-pressure traffic signal controller based on more realistic assumptions on the available measurements than state-of-the-art back-pressure traffic signal control and to show in simulations that stability is conserved in a significant part of the capacity region.

The paper is organized as follows. Section \ref{sec-model} describes the queuing network model. Sections \ref{sec-bp-control} is mainly expository:  it describes BP* highlighting its stability-optimality. The contributions of the paper are presented in Section \ref{sec-bp-aggregated} and \ref{sec-simulations}. Section \ref{sec-bp-aggregated} exhibits BP and a theoretical result on the Lyapunov drift that tends to indicate that it should have good stability properties. The simulations of Section \ref{sec-simulations} confirm this and show that BP stabilizes the network in a significant part of the capacity region. Section \ref{sec-conclusion} concludes the paper and opens perspectives.

\section{Model}
\label{sec-model}

As standard in queuing network control, time is slotted, and each time slot maps to a certain period of time during which a control is applied. It is convenient to use a fixed pre-defined time slot length, whose size corresponds to the minimal duration of a phase. When the time slot size is fixed, the traffic signal control problem consists of computing at the beginning of each time slot $t$ the phase to apply during slot $t$. The network of intersections is modelled as a directed graph of nodes $(\node_a)_{a\in\nodeset}$ and links $(\link_j)_{j\in\linkset}$. Nodes represent lanes with queuing vehicles, and links enable transfers from node to node: this is a standard queuing network model.

\begin{figure}[ht]
\centering
\includegraphics[width=0.8\linewidth]{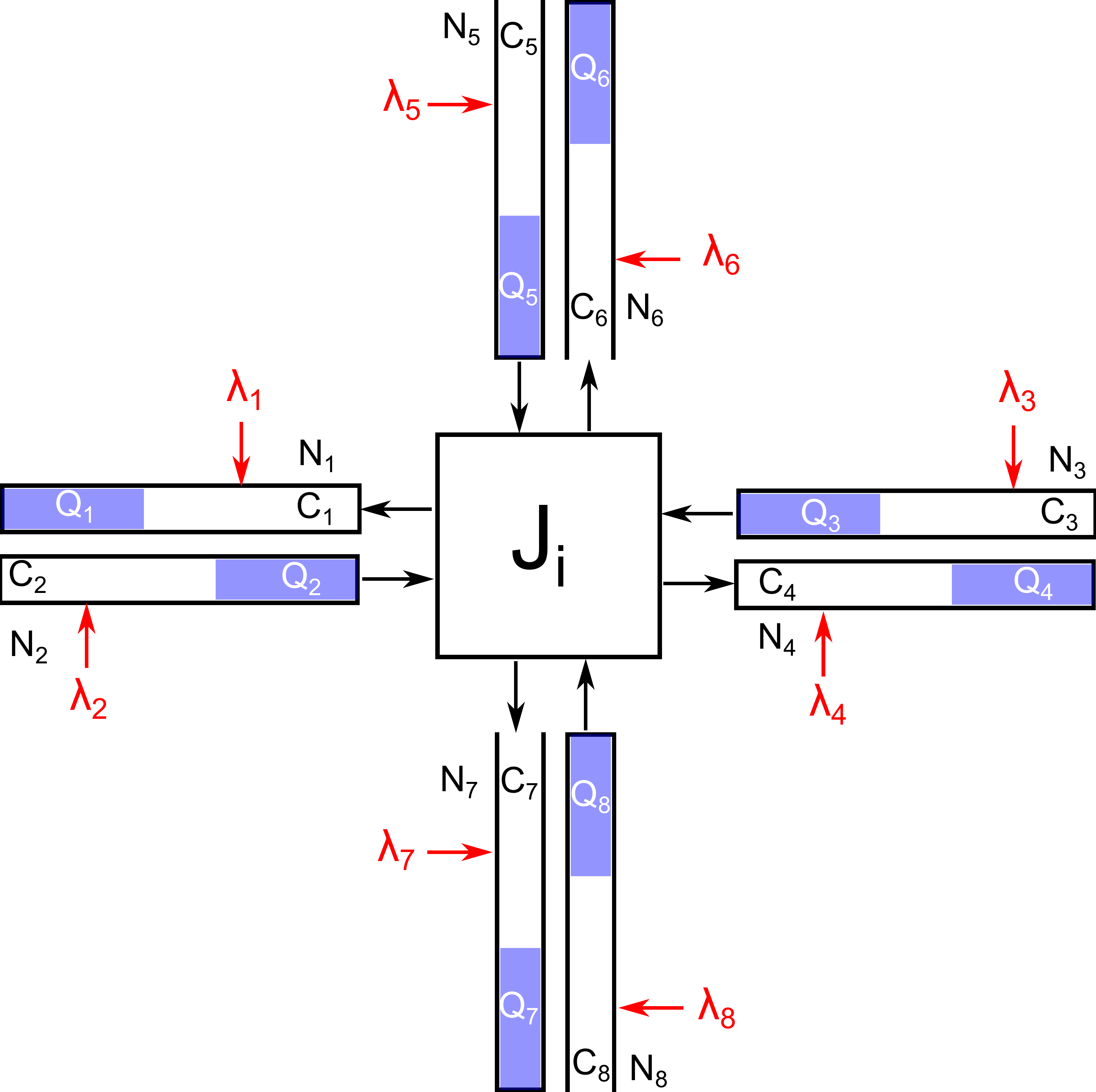}\hfill
\caption{A junction with 4 incoming nodes and 4 outgoing nodes which corresponds to the intersection depicted in Figure \ref{fig-phases}.}
\label{fig-network-topology}
\end{figure}

It is a multiple queues one server queuing network. Every signalized intersection is modelled as a server managing a junction which consists of set of links. Junctions $(\junction_i)_{i\in\junctionset}$ are supposed to form a partition of links. For every junction $\junction$, $\inputnodes(\junction)$ and $\outputnodes(\junction)$ denote respectively the inputs and the outputs of $\junction$. Inputs (resp. outputs) of junction $\junction$ are nodes $N$ such that there exists a link $L\in\junction$ pointing from (resp. to) $N$. The reader should consider the introduction of junctions in the model as an overlay of the queuing network model. For the sake of simplicity, we do not represent links in the queuing network representation of Figure \ref{fig-network-topology}.

Every server maintains an internal queue for every input/output, and server work enables to transfer vehicles from an input to an output of the junction. The internal queue at node $\node_a$ is a vector $Q_{a}$ and $Q_{ab}(t)$ denotes the number of vehicles in the queue of node $\node_a$ entering $\node_b$ upon leaving $\node_a$. The aggregated queue length $Q_a(t)=\sum_b Q_{ab}(t)$ denotes the total number of vehicles at node $\node_a$ considering all possible routings after exiting $\node_a$. In this paper, queues are supposed to have infinite capacities: there is no blocking (see 
\cite{gregoire2013capacity} for an adaptation of back-pressure traffic signal control in the context of finite capacities).

\begin{figure}[ht]
\centering
\includegraphics[width=0.8\linewidth]{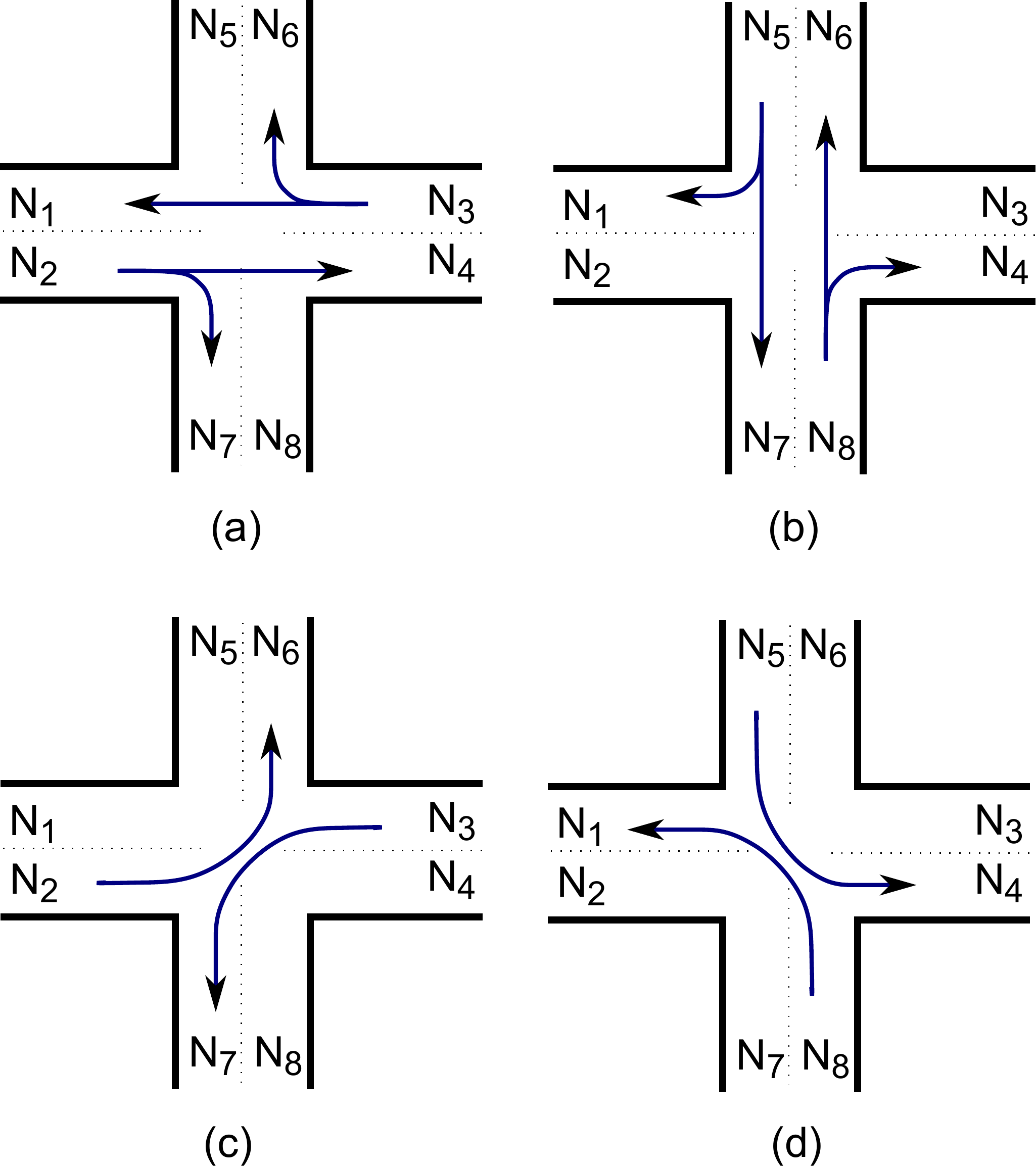}\hfill
\caption{A typical set of feasible phases at a junction. For example, supposing that service rates equal $0$ or $1$, the non zero service rates for phase (a) are $\mu_{31}$, $\mu_{36}$, $\mu_{24}$ and $\mu_{27}$}
\label{fig-phases}
\end{figure}

At every time slot $t$, servers work, resulting in vehicles transfers. Under phase-based control, the transmission rate offered by servers are set by activating a given signal phase $p_i$ at each junction $\junction_i$ from a predefined finite set of feasible phases $\phases_i$ at every time slot $t$. Let $\phases=\prod_{i\in\junctionset} \phases_i$ denote the set of feasible global phases. Each global phase $p=(p_i)_{i\in\junctionset}\in\phases$ results in a different service matrix $\mu(p)$ where $\mu_{ab}(p)$ represents the transmission rate offered by servers to transfer vehicles from $\node_a$ to $\node_b$ in a time slot when phase $p$ is activated. The transmission rate is assumed to be binary, $\mu_{ab}(p)\in\{0,s_{ab}\}$: it is zero or it equals the saturation rate $s_{ab}$. Only the vehicles which are at a  node at the beginning of time slot $t$ can be transferred from that node to another node during slot $t$. Figure \ref{fig-phases} depicts the 4 typical phases of a 4 inputs/4 outputs junction.

Exogenous arrivals occur at every node of the network. Let $A_a(t)$ denote the number of vehicles that exogenously arrive at node $\node_a$ during slot $t$. The arrival process $A_a(t)$ is assumed to be rate convergent with long-term arrival rate $\lambda_a \geq 0$. When a quantity of vehicles arrives at node $N_a\in\inputnodes(\junction_i)$ during slot $t$, exogenously and endogenously, it is split and added into queues $Q_{ab}$, $b\in\outputnodes(\junction_i)$. The routing process is exogenous and assumed to be rate convergent with ratios $r_{ab}$ with $\sum_b r_{ab} \leq 1$ (see the supplementary material \cite{gregoire2013supplementBackpressure} for more details). Exits are modelled by assuming that the routing matrix is non-conservative. $1-\sum_b r_{ab}$ represents the exit rate of vehicles entering node $\node_a$, it is the ratio of vehicles directly removed from the network when entering node $\node_a$, i.e. not added to any queue $Q_{ab}$. Note that the only variable that is controlled is the activated phase at every time slot $t$, denoted by $p(t)$, and yielding a service matrix $\mu(p(t))$ during slot $t$.

\section{BP* controller}
\label{sec-bp-control}

\subsection{The controller}

In the following, we expose BP* signal control. It is an extension of the algorithm proposed in \cite{Varaiya2009} where internal/exit links are not differentiated, because exits may occur at any link of the network. It is quite equivalent to the back-pressure controller of \cite{Wongpiromsarn2012}, assuming the nodes carry direction information. Loosely speaking, the idea of back-pressure control is to compute pressure at every node based on node occupancy and to open flows which have a high upstream pressure and a low downstream pressure, like opening a tap. 

\begin{algorithm}[ht]
\begin{algorithmic}[5]
\Require 
\\Queues lengths matrix $Q(t)$,
\\ Pressure functions $P_{ab}(Q_{ab})$ for all $a,b\in\nodeset$,
\\ Routing matrix $r$.
\Function{BP*}{}
\For{$i \in \junctionset$}
\For{$a\in\inputnodes(\junction_i),b\in\outputnodes(\junction_i)$}
\State $\Pi_{ab}(t) \gets P_{ab}\left[Q_{ab}(t)\right]$
\EndFor
\For{$a\in\inputnodes(\junction_i),b\in\outputnodes(\junction_i)$}
\State $W_{ab}(t) \gets \max\left(\Pi_{ab}(t)-\sum_c r_{bc} \Pi_{bc}(t),0\right)$
\EndFor
\State $p^\star_i(t) \gets \arg \max\limits_{p_i\in\phases_i} \sum\limits_{a\in\inputnodes(\junction_i),b\in\outputnodes(\junction_i)} W_{ab}(t) \mu_{ab}(p_i)$ \label{line-arg-max-bp}
\EndFor
\State \Return{Phase $p^\star(t)$ to apply in time slot $t$}
\EndFunction
\end{algorithmic}
\caption{BP* control}
\label{alg-back-pressure}
\end{algorithm}

Algorithm \ref{alg-back-pressure} defines BP* control. At every junction $i$, for each phase $p\in\phases_i$, the weighted sum $\sum_{a,b} W_{ab}(t) \mu_{ab}(p)$ is computed. $W_{ab}(t)$, the weight associated to transfers from $\node_a$ to $\node_b$, is the difference between the upstream pressure $\Pi_{ab}(t)$ and the weighted downstream pressure $\sum_c r_{bc}\Pi_{bc}(t)$. BP* consists of selecting the phase that maximizes the weighted sum. Moreover, we assume that in case of equality the selected phase $p^*(t)$ always satisfies $\mu_{ab}(p^*(t))=0$ if $W_{ab}(t)=0$. 

\subsection{Optimal stability}

The following theorem states that under linear pressure functions with strictly positive slope, BP* as defined by Algorithm \ref{alg-back-pressure} is optimal in terms of stability, i.e. stabilizes the network for all arrivals rates that can be stabilized considering all possible control strategies. It is an extension of the results of \cite{Varaiya2009}, because vehicles can enter/exit the network at any node, there is no distinction between exit nodes and internal nodes. Moreover, in contrast with \cite{Varaiya2009, Wongpiromsarn2012}, pressure functions are just assumed to be linear with strictly positive slope in this paper: $P_{ab}(Q_{ab})=\theta_{ab} Q_{ab}$, $\theta_{ab}>0$.

\begin{thm}[Back-pressure optimality]
\label{thm-bp-optimality}
Assuming that \\pressure functions are linear with strictly positive slopes, BP* as defined by Algorithm \ref{alg-back-pressure} is stability-optimal.
\end{thm}

\begin{pf}
Due to space limitations, the full proof is not provided in this paper and is available in the supplementary material \cite{gregoire2013supplementBackpressure}. Stability is proved using the Lyapunov function $V(t)=\mathcal{V}(Q(t)) = \sum_{a,b} \theta_{ab} Q_{ab}(t)^2$. The existence of $B,\eta>0$ such that: 

\begin{multline}
\expect\{V(t+1)-V(t) | Q(t)\} \leq B - \eta \sum_{a,b} Q_{ab}(t)\text{,}
\end{multline}
enables to conclude stability for the queuing network using the sufficient condition  proved in \cite{Neely2003}.
\end{pf}

\section{BP controller}
\label{sec-bp-aggregated}

\subsection{The controller}

Back-pressure control proposed in Section \ref{sec-bp-control} requires complete knowledge of the queues lengths matrix $Q(t)$ and the routing rates. For our application, a complete knowledge of $Q(t)$ is not realistic because dedicated lanes for turning vehicles are only from the proximity of the junction. Farther, all vehicles are gathered and the controller does not have access to the direction of every vehicle in the absence of vehicle-to-infrastructure communications. That is why we propose in the present paper a controller that uses only the aggregated queues lengths $Q_a(t)=\sum_b Q_{ab}(t)$, i.e. a queue length without direction information. It is defined by Algorithm \ref{alg-back-pressure-aggregated}. It computes the phase to apply at every time slot without requiring neither routing rates nor complete knowledge of queues lengths matrix $Q(t)$ and takes as inputs the aggregated queues lengths ${Q_a(t) = \sum_b Q_{ab}(t)}$. However, it still requires vehicle detectors variables ${d_{ab}(t)\in[0,1]}$ defined below:

\begin{equation}
d_{ab}(t) = \min(Q_{ab}(t)/s_{ab},1)
\end{equation}

The variable $d_{ab}(t)$ can be measured by loop detectors positioned at dedicated lanes.

\begin{algorithm}[ht]
\begin{algorithmic}[5]
\Require 
\\Queues lengths $Q_a(t)$,
\\ Pressure functions $P_{a}(Q_{a})$,
\\ Loop detectors variables $d_{ab}(t)$.
\Function{BP}{}
\For{$i \in \junctionset$}
\For{$a\in\inputnodes(\junction_i) \cup \outputnodes(\junction_i)$}
\State $\Pi_{a}(t) \gets P_{a}\left[Q_{a}(t)\right]$
\EndFor
\For{$a\in\inputnodes(\junction_i),b\in\outputnodes(\junction_i)$}
\State $W_{ab}(t) \gets d_{ab}(t) \max\left(\Pi_{a}(t)-\Pi_{b}(t),0\right)$
\EndFor
\State $p^\star_i(t) \gets \arg \max\limits_{p_i\in\phases_i} \sum\limits_{a\in\inputnodes(\junction_i),b\in\outputnodes(\junction_i)} W_{ab}(t) \mu_{ab}(p_i)$ \label{line-arg-max-bp-aggregated}
\EndFor
\State \Return{Phase $p^\star(t)$ to apply in time slot $t$}
\EndFunction
\end{algorithmic}
\caption{BP control}
\label{alg-back-pressure-aggregated}
\end{algorithm}

Algorithm \ref{alg-back-pressure-aggregated} defines BP control. Note that for transfers from $\node_a$ to $\node_b$, the upstream pressure is now $\Pi_a(t)$ and the downstream pressure is $\Pi_b(t)$: individual queue pressures $\Pi_{ab}(t)$ are not required. Moreover, the difference between the upstream pressure and the downstream pressure is multiplied by $d_{ab}(t)$ to form $W_{ab}(t)$. Hence, if at time slot $t$, there is no vehicle at $\node_a$ going to $\node_b$, the weight $W_{ab}(t)$ associated to transfers from $\node_a$ to $\node_b$ vanishes.

\subsection{Behaviour of the Lyapunov drift under heavy load conditions}

Let us consider the Lyapunov function $\mathcal{V}(Q)$ and its evolution through time $V(t)$ defined below:

\begin{equation}
V(t)=\mathcal{V}(Q(t))=\sum_a \theta_a Q_a(t)^2 = \sum_a \theta_a (\sum_b Q_{ab}(t))^2
\end{equation}

Let us define heavy load conditions at time slot $t$ as states of the network such that if the right-of-way is given to any individual queue, it can be emptied at saturation flow, i.e. there are enough vehicles in the individual queue to ensure saturation:

\begin{equation}
\forall a,b\in\nodeset, Q_{ab}(t)\geq s_{ab}
\end{equation}

The following theorem proves that under heavy load conditions the Lyapunov drift respects the sufficient condition for network stability if $\lambda+\epsilon \in\Lambda_r$, for sufficiently large $\epsilon$. 

\begin{thm}[Lyapunov drift under heavy load conditions]
\label{thm-lyapunov-drift}
Assume $\lambda+\epsilon\in\Lambda_r$, BP control as defined in Algorithm \ref{alg-back-pressure-aggregated} is applied and the network is in heavy load conditions, then there exists $B,\eta>0$ such that :

\begin{equation}
\expect\{ V(t+1) - V(t)~|~Q(t)\} \leq B - \eta \sum_a Q_a(t)
\end{equation}
for sufficiently large $\epsilon$.
\end{thm}

\begin{pf}
Due to space limitations, the full proof is not provided in this paper and is available in the supplementary material \cite{gregoire2013supplementBackpressure}. 
\end{pf}

The above theorem tends to indicate that the network should have good stability properties because the condition for stability is verified in heavy load conditions for $\lambda$ sufficiently interior to the capacity region. Unfortunately it does not enable to conclude that the network is stable in a significant part of the capacity region. Indeed, heavy load conditions can not be guaranteed at all time, and when an individual queue $Q_{ab}$ is below the saturation flow $s_{ab}$, it is a constraint for the emptying of $Q_a$, that can unstabilize the queuing network. Hence, the characterization of the stability region of the queuing network under BP control with the modelling assumptions presented in Section \ref{sec-model} is still a challenging problem. That is why we propose to implement the two back-pressure controllers and to compare their behaviour. The results of the simulations are presented in the next section.

\section{Simulations}
\label{sec-simulations}

\subsection{The simulation platform}

The model and the algorithms presented in this paper have been implemented into a simulator coded in Java. It simulates a grid network and every junction of the grid has 4 inputs, 4 outputs, and 4 feasible phases as depicted in Figure \ref{fig-phases}. The height and the width are parameters. Every individual flow allowed by phases of Figure \ref{fig-phases} equals $10$ (it is the saturation rate). Vehicles are generated at each node $\node_a$ at an arrival rate $\lambda_a$ that can be set as desired. The arrival process generates individual arrivals as well as batches of 10 vehicles. The routing ratios are fixed at the beginning of the simulation.

\subsection{Behaviour of the two back-pressure controllers}

\begin{figure}[ht]
\centering
\includegraphics[width=0.6\linewidth]{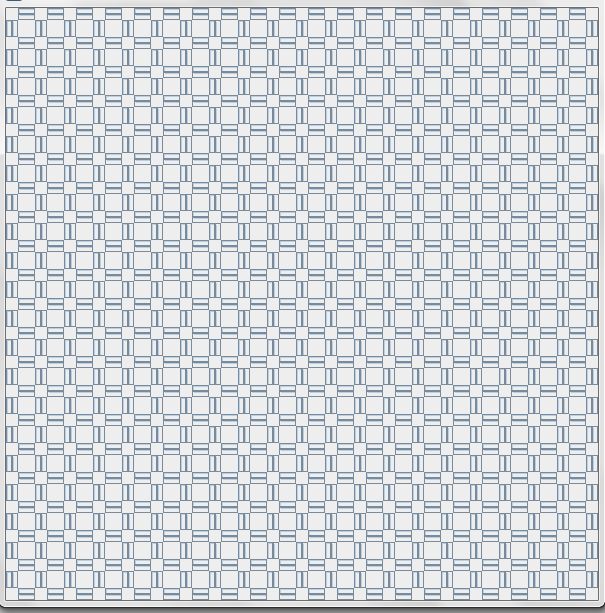}\hfill
\caption{The $21 \times 21$ grid network used for the presented simulations.}
\label{fig-grid-network}
\end{figure}

Simulations have been carried out for a $21 \times 21$ square grid network (see Figure \ref{fig-grid-network}). First of all, we present simulations results in the case of a network that has been configured with the same arrival rates and routing rates at every node of the network. 

\subsubsection{Simulation results for a particular network and particular arrival/routing rates}

\begin{figure}[ht]
\centering
\includegraphics[width=0.5\linewidth]{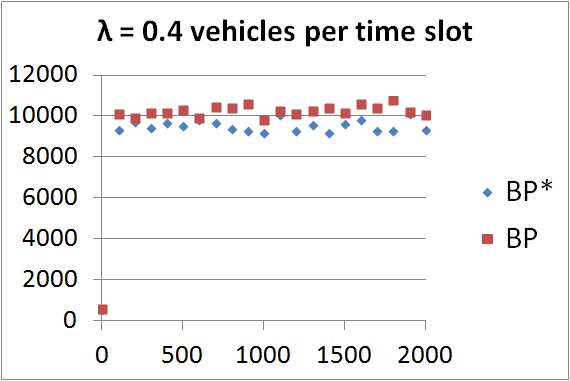}\hfill
\includegraphics[width=0.5\linewidth]{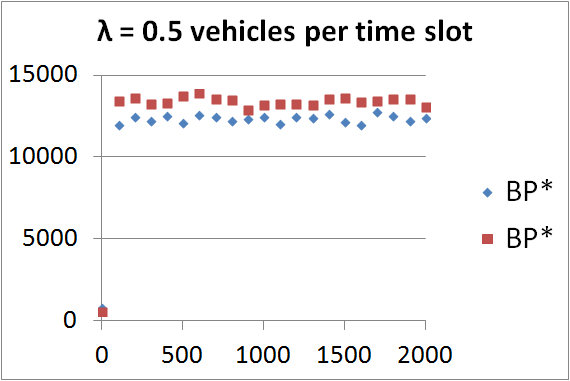}\hfill
\includegraphics[width=0.5\linewidth]{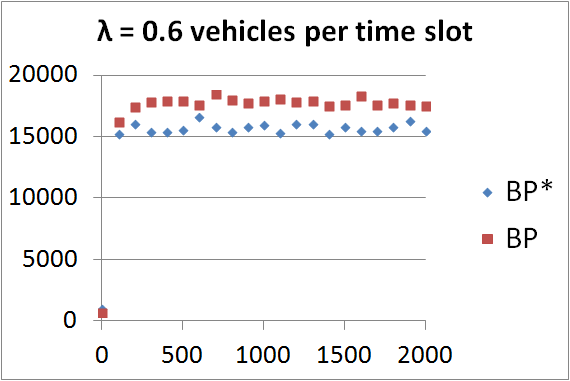}\hfill
\includegraphics[width=0.5\linewidth]{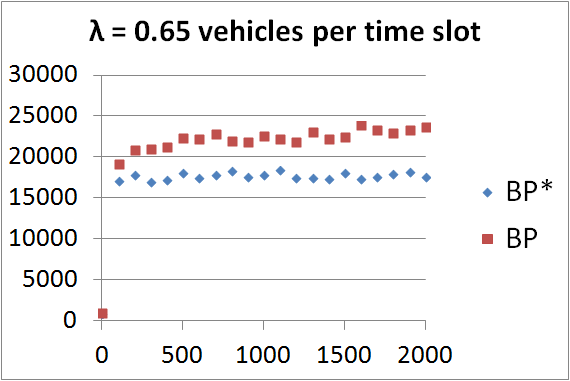}\hfill
\includegraphics[width=0.5\linewidth]{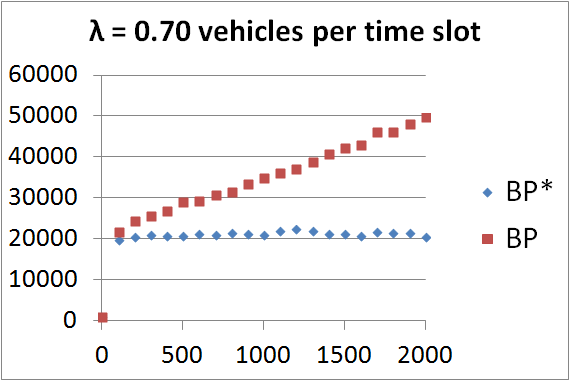}\hfill
\includegraphics[width=0.5\linewidth]{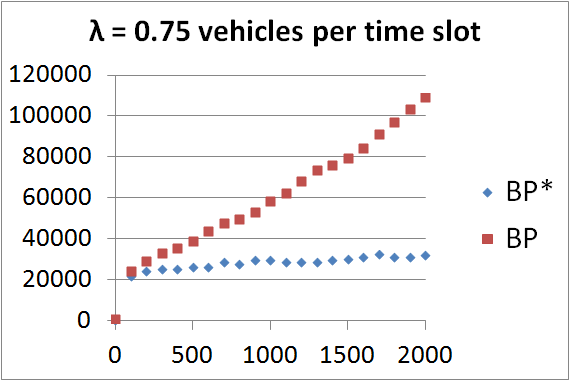}\hfill
\includegraphics[width=0.5\linewidth]{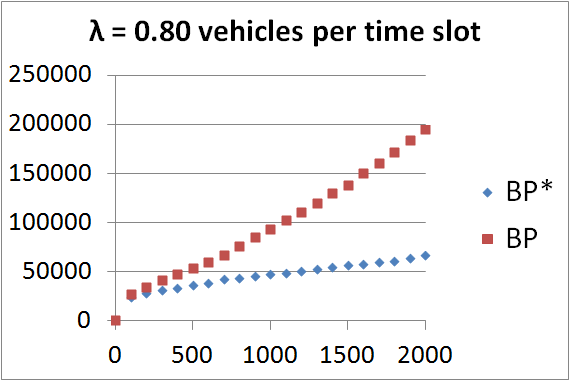}\hfill
\includegraphics[width=0.5\linewidth]{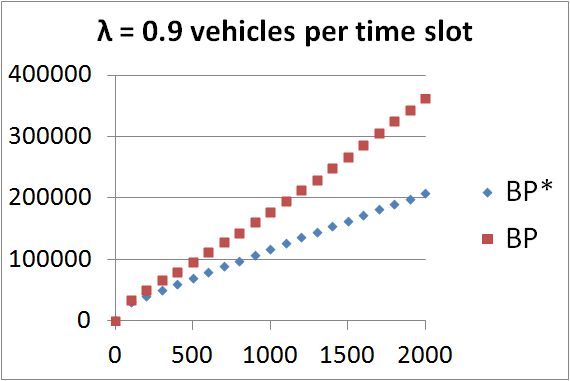}\hfill
\caption{Evolution of the global queue of the network over time for height arrival rates. Comparison of the behaviour of the network under BP/BP* control.}
\label{fig-simulation-results}
\end{figure}

The numerical results of Figure \ref{fig-simulation-results} correspond to the following parameters. Turn left probability when a vehicle enters a node: $0.2$; turn right probability when a vehicle enters a node: $0.2$; go straight probability when a vehicle enters a node: $0.5$; exit probability when a vehicle enters a node: $0.1$; probability of a batch: $0.05$; pressure functions $P_a(Q_a)=Q_a$ and $P_{ab}(Q_{ab})=Q_{ab}$ ($\theta_a=\theta_{ab}=1$); vehicles are generated at every node with the same arrival rate $\lambda>0$ that can be set as desired at the beginning of the simulation.

Experiments are carried out at height different arrival rates: $\lambda = $ 0.4, 0.5, 0.6, 0.65, 0.7, 0.75, 0.8 and 0.9 vehicles per time slot. Figure \ref{fig-simulation-results} depicts the global queue of the network over time, i.e. $\sum_a Q_a(t)=\sum_{a,b} Q_{ab}(t)$, for the height arrival rates, under BP* control and under BP control. One can observe in Figure \ref{fig-simulation-results} that under BP* control, the queuing network is stabilized for $\lambda \leq 0.7$ and gets unstable from $\lambda=0.75$. Under BP control, it is stabilized for $\lambda \leq 0.65$ and gets unstable from $\lambda = 0.7$. First of all, it proves that as expected, BP control is not stability-optimal.  However, in the particular setting of this experiment, (uniform arrivals/routing rates and grid network), the performance of BP and BP* are very close, and the optimality gap is around $0.05/0.7 \simeq 10\%$, i.e. a performance of 90\%. 

However, such a uniform network is not realistic and the results of the next paragraph try to evaluate the performance of BP with regards to BP* with less specific routing/arrival parameters. 

\subsubsection{Evaluation of BP with regards to BP* on several samples of parameters}

In the following simulations, the routing/arrival process parameters are not uniform over nodes any more. 10 samples of parameters have been generated. For each sample, the routing/arrival rates are generated as follows. For each direction (straight, left, right), (uniformly) random values between 0 and 1 are generated, say $y_s,y_l,y_r$; a (uniformly) random value between 0 and 0.1 is generated for exits, say $y_\omega$; and the routing rates are set by normalization of the generated real values, i.e. for the left direction for example, the routing rate is $y_l/(y_s+y_l+y_r+y_\omega)$. The arrivals rates are set by generating a (uniformly) random value between 0 and 1 for every node, say $\lambda_a^0$. At the beginning of the simulation, a parametrizable scaling value $x$ enables to fix the actual arrival rate of the current simulation: $\lambda_a = x \lambda_a^0$, where $x$ has the same value over nodes. The value of 0.1 for the scale of exits is quite arbitrary and, loosely speaking, fixes the averaged number of travelled nodes before exiting the network. 

Note that the routing rates and the values $\lambda_a^0$ are fixed for a given sample. However, the value of $\lambda_a$ depends on the value of $x$ set at the beginning of the simulation. The parameter $x$ enables to define a performance for BP with regards to BP* for a given sample. We let $x$ vary and we observe the maximum value of $x$ such that the network is stable under BP versus BP* (say $x_\mathrm{\max}^{\mathrm{*}}$ for BP* and $x_\mathrm{\max}^{\mathrm{BP}}$ for BP). We define the performance of BP with regards to BP*, or more shortly the performance of BP (because BP* is optimal), as follows:

\begin{equation}
\mathrm{performance(BP)} = x_\mathrm{\max}^{\mathrm{BP}}/x_\mathrm{\max}^{\mathrm{*}}
\end{equation}

\begin{figure}[ht]
\centering
\includegraphics[width=1\linewidth]{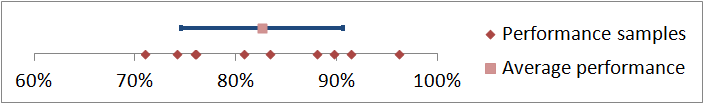}\hfill
\caption{Performance distribution for ten samples. The point above the axis represents the average performance over samples and the horizontal bar is the standard deviation.}
\label{fig-performance-samples}
\end{figure}

As for previously presented simulations, the probability of a batch is 0.05 and the pressure functions are linear with slope 1. Figure \ref{fig-performance-samples} depicts the performance obtained for the 10 samples, the average performance and the standard deviation. The average performance is around 80\%, i.e. the optimality gap is about 20\%. The simulation results prove that the performance of BP is affected by the routing/arrival rates. Hence, the distribution (over samples) of the performance would be different for a different distribution of routing/arrival rates. Nevertheless, in the particular setting of the experiment, the average optimality gap of 20\% seems again a low price to pay with regards to the much more realistic assumptions on the measurements available to compute the control.

However, these promising results can not be extended to any kind of network of intersections and further simulations with a more general structure of network should be carried out to confirm the closeness of performance. We are currently implementing our algorithms in a traffic simulator in order to test the performance of BP control with real traffic data of the city of Singapore.

\section{Conclusion and perspectives}
\label{sec-conclusion}

The simulation results of this paper prove that BP is not optimal but tend to indicate that it stabilizes the queuing network in a significant part of the capacity region. The benefits of BP originate from the more realistic assumptions on queues measurements. Computing the phase to apply only requires aggregated queues lengths estimation that can be provided by cameras, and loop detectors at dedicated lanes. The optimality gap, around 20\% in the particular setting of the experiments, seems a low price to pay for the benefits of relaxed assumptions on the available measurements. However, simulations have been conducted in a grid network, which is a particular structure, and with synthetic data which can strongly differ from real traffic data. To confirm the closeness of performance, simulations should be carried out in a more advanced traffic network simulator. 

Finally, the emergence of vehicle-to-infrastructure communications opens avenues to enhance traffic signal control. The traffic signal controllers can have access, in particular, to the destination node of every vehicle. As a result, back-pressure control with a multiple-commodity queuing network model, as proposed in \cite{Neely2003} in the context of wireless communication networks, should be investigated.

\begin{ack}
This work was supported in part by the Singapore National Research Foundation through the Future Urban Mobility Interdisciplinary Research Group at the Singapore-MIT Alliance for Research and Technology.
\end{ack}

\bibliographystyle{ifacconf}
\bibliography{biblio}

\begin{thebibliography}{18}
\providecommand{\natexlab}[1]{#1}
\providecommand{\url}[1]{\texttt{#1}}
\providecommand{\urlprefix}{URL }
\expandafter\ifx\csname urlstyle\endcsname\relax
  \providecommand{\doi}[1]{doi:\discretionary{}{}{}#1}\else
  \providecommand{\doi}{doi:\discretionary{}{}{}\begingroup
  \urlstyle{rm}\Url}\fi

\bibitem[{Cascetta et~al.(2006)Cascetta, Gallo, and
  Montella}]{cascetta2006models}
Cascetta, E., Gallo, M., and Montella, B. (2006).
\newblock Models and algorithms for the optimization of signal settings on
  urban networks with stochastic assignment models.
\newblock \emph{Annals of Operations Research}, 144(1), 301--328.

\bibitem[{Diakaki et~al.(2002)Diakaki, Papageorgiou, and
  Aboudolas}]{Diakaki2002}
Diakaki, C., Papageorgiou, M., and Aboudolas, K. (2002).
\newblock A multivariable regulator approach to traffic-responsive network-wide
  signal control.
\newblock \emph{Control Engineering Practice}, 10(2), 183--195.

\bibitem[{Gartner(1983)}]{Gartner1983}
Gartner, N.H. (1983).
\newblock Opac: A demand-responsive strategy for traffic signal control.
\newblock \emph{Transportation Research Record}, (906).

\bibitem[{Gartner et~al.(1975)Gartner, Little, and
  Gabbay}]{gartner1975optimization}
Gartner, N.H., Little, J.D., and Gabbay, H. (1975).
\newblock Optimization of traffic signal settings by mixed-integer linear
  programming part i: The network coordination problem.
\newblock \emph{Transportation Science}, 9(4), 321--343.

\bibitem[{Gregoire et~al.(2013{\natexlab{a}})Gregoire, Frazzoli,
  de~La~Fortelle, and Wongpiromsarn}]{gregoire2013capacity}
Gregoire, J., Frazzoli, E., de~La~Fortelle, A., and Wongpiromsarn, T.
  (2013{\natexlab{a}}).
\newblock Capacity-aware back-pressure traffic signal control.
\newblock \emph{arXiv preprint arXiv:1309.6484}.

\bibitem[{Gregoire et~al.(2013{\natexlab{b}})Gregoire, Frazzoli,
  de~La~Fortelle, and Wongpiromsarn}]{gregoire2013supplementBackpressure}
Gregoire, J., Frazzoli, E., de~La~Fortelle, A., and Wongpiromsarn, T.
  (2013{\natexlab{b}}).
\newblock Supplementary material to: Back-pressure traffic signal control with
  unknown routing rates.
\newblock \emph{hal preprint hal-00905063}.
\newblock \urlprefix\url{http://hal.archives-ouvertes.fr/hal-00905063}.

\bibitem[{Henry et~al.(1984)Henry, Farges, and Tuffal}]{Henry1984}
Henry, J.J., Farges, J.L., and Tuffal, J. (1984).
\newblock The prodyn real time traffic algorithm.
\newblock In \emph{Proceedings of the 4th IFAC/IFORS Conference on Control in
  Transportation Systems,}.

\bibitem[{Hunt et~al.(1982)Hunt, Robertson, Bretherton, and Royle}]{Hunt1982}
Hunt, P., Robertson, D., Bretherton, R., and Royle, M. (1982).
\newblock The scoot on-line traffic signal optimisation technique.
\newblock \emph{Traffic Engineering \& Control}, 23(4).

\bibitem[{Lowrie(1990)}]{Lowrie1990}
Lowrie, P. (1990).
\newblock Scats, sydney co-ordinated adaptive traffic system: A traffic
  responsive method of controlling urban traffic.
\newblock Technical report.

\bibitem[{Miller(1963)}]{miller1963settings}
Miller, A.J. (1963).
\newblock Settings for fixed-cycle traffic signals.
\newblock \emph{Operations Research}, 373--386.

\bibitem[{Mirchandani and Head(2001)}]{Mirchandani2001}
Mirchandani, P. and Head, L. (2001).
\newblock A real-time traffic signal control system: architecture, algorithms,
  and analysis.
\newblock \emph{Transportation Research Part C: Emerging Technologies}, 9(6),
  415--432.

\bibitem[{Neely(2003)}]{Neely2003}
Neely, M.J. (2003).
\newblock \emph{Dynamic power allocation and routing for satellite and wireless
  networks with time varying channels}.
\newblock Ph.D. thesis, LIDS, Massachusetts Institute of Technology.

\bibitem[{Papageorgiou et~al.(2003)Papageorgiou, Diakaki, Dinopoulou,
  Kotsialos, and Wang}]{Papageorgiou2003}
Papageorgiou, M., Diakaki, C., Dinopoulou, V., Kotsialos, A., and Wang, Y.
  (2003).
\newblock Review of road traffic control strategies.
\newblock \emph{Proceedings of the IEEE}, 91(12), 2043--2067.

\bibitem[{Shepherd(1992)}]{Shepherd1992}
Shepherd, S. (1992).
\newblock A review of traffic signal control.
\newblock Technical report.

\bibitem[{Tassiulas and Ephremides(1992)}]{Tassiulas1992}
Tassiulas, L. and Ephremides, A. (1992).
\newblock Stability properties of constrained queueing systems and scheduling
  policies for maximum throughput in multihop radio networks.
\newblock \emph{IEEE Transactions on Automatic Control}, 37(12), 1936--1948.

\bibitem[{Varaiya(2009)}]{Varaiya2009}
Varaiya, P. (2009).
\newblock A universal feedback control policy for arbitrary networks of
  signalized intersections.
\newblock Technical report.

\bibitem[{Varaiya(2013)}]{Varaiya2013}
Varaiya, P. (2013).
\newblock The max-pressure controller for arbitrary networks of signalized
  intersections.
\newblock In \emph{Advances in Dynamic Network Modeling in Complex
  Transportation Systems}, 27--66. Springer.

\bibitem[{Wongpiromsarn et~al.(2012)Wongpiromsarn, Uthaicharoenpong, Wang,
  Frazzoli, and Wang}]{Wongpiromsarn2012}
Wongpiromsarn, T., Uthaicharoenpong, T., Wang, Y., Frazzoli, E., and Wang, D.
  (2012).
\newblock Distributed traffic signal control for maximum network throughput.
\newblock In \emph{Proceedings of the 15th international IEEE conference on
  intelligent transportation systems}, 588--595.

\end{thebibliography}

\end{document}